\documentclass[]{aastex}

\usepackage{epsf}
\usepackage[dvips]{epsfig}

\begin{document}

\title{The possible origin of the faint fuzzy star clusters in
  NGC~1023} 
\author{M. Fellhauer, P. Kroupa}
\affil{Institute for Theoretical Physics and Astrophysics,
  University of Kiel, Germany}
\email{mike,pavel@astrophyik.uni-kiel.de}

\begin{abstract}
In the lenticular galaxy NGC~1023 a new population of star
clusters (``faint fuzzies'') was recently discovered by Larsen \&
Brodie.  These clusters are found inside the disc and are faint
($23 \leq V \leq 24$~mag) and extended with effective radii of
$r_{\rm eff} \approx 7$ to $15$~pc.  \\
We present here N-body calculations of a likely
formation-scenario through merging star clusters in clusters of
star clusters (super-clusters).  Such super-clusters are observed
to form in interacting galaxies.  \\
The resulting merger objects have masses comparable to the
``faint fuzzies'' and show large effective radii ($r_{\rm eff} >
7$~pc).  Even though these objects are suffering from strong
tidal forces they are able to survive and reach the estimated
ages of the extended star clusters in NGC~1023. 
\end{abstract}


\maketitle

\section{Introduction}
\label{sec:intro}

During a recent search for globular clusters (GC) in NGC~1023, a 
lenticular galaxy at approximately $9$~Mpc distance, Larsen \&
Brodie (2000) found faint objects ($23 \leq V \leq 24$~mag) that
have GC luminosities but effective radii ($=$ half-light radii)
of $7$ to $15$~pc which is much larger than ordinary GCs that
have effective radii of $2$ to $3$~pc.  Using spectra obtained
with the Keck telescope they (Brodie \& Larsen 2002)
confirmed that these objects belong to NGC~1023 and have ages of
about $\geq 7$ to $8$~Gyr, are moderately metal rich ($[\rm Fe/H]
= -0.58 \pm 0.24$) while showing a super-solar $\alpha$-element
abundance ($[\alpha/{\rm Fe}] = +0.3$ to $+0.6$).  While GC are
usually connected with the spherical component of a galaxy, these
new objects appear to be situated in the disc of NGC~1023.  The
distribution of these objects shows rotation around the galaxy.
The authors conclude that they have found a new class of star
clusters which they call ``faint fuzzies''. \\ 

NGC~1023 has a dwarf companion (NGC~1023A) with which it is
steadily interacting, and due to its lenticular appearance it
probably suffered from interactions and mergers in the past
leading to strong star formation events. \\ 

Interactions between gas-rich galaxies lead to strong star
formation events.  For example, HST-images of the Antennae reveal
(Whitmore et al.\ 1999) that the knots of intense star formation
produce clusters of massive young star clusters.  These
``super-clusters'' ($=$ clusters of star clusters; not to be
confused with super stellar clusters (SSC), which are individual
massive star clusters) appear to contain dozens to hundreds of
massive star clusters within a region spanning a few hundred pc
to a kpc in diameter. \\  

Here we report N-body results concerning the dynamical evolution
of such super-clusters.  The resulting merger objects have
similar properties as the ``faint fuzzies''.

\section{Simulations}
\label{sec:setup}

We use the particle-mesh code {\sc Superbox} (Fellhauer et al.\
2000) which incorporates a hierarchical grid architecture
allowing high resolution at the places of interest. \\

We model the single star clusters as Plummer spheres (Plummer
1911, numerical realisation: Aarseth et al.\ 1974) with a Plummer
radius of $4$~pc, which corresponds to the mean half-light radius
found for the individual young star clusters in the Antennae
(Whitmore et al.\ 1999).  $N_{0} = 262$ star clusters with a mass
range of $10^{4} - 10^{6}$~M$_{\odot}$ following a power-law mass
spectrum, 
\begin{eqnarray}
  \label{eq:massspec}
  n(M_{\rm cl}) & \propto & M_{\rm cl}^{-1.5},
\end{eqnarray}
are placed in a Plummer distribution with Plummer radius $r_{\rm
  pl}^{\rm sc}$ representing the super-cluster.  The
super-cluster orbits on an eccentric orbit around the host galaxy
represented by an analytical potential, which consists of a disc
modelled as a Plummer-Kuzmin potential and a spherical halo
component modelled as a logarithmic potential:
\begin{eqnarray}
  \label{eq:pot}
  \Phi_{\rm gal} & = & \Phi_{\rm disc} + \Phi_{\rm halo} \\
  & = & 
- \frac{GM_{\rm disc}}{\sqrt{R^{2} + ( a +
      \sqrt{z^{2}+b^{2}})^{2} } } - \frac{1}{2} v_{0}^{2} \ln(
  R_{\rm gal}^{2} + R^{2}), \nonumber
\end{eqnarray}
with $M_{\rm disc} = 10^{11}$~M$_{\odot}$, $a = 3$~kpc, $b =
0.3$~kpc, $v_{0} = 200$~km/s and $R_{\rm gal} = 50$~kpc which
sums up to an almost flat rotation curve with a rotation speed of
$220$~kms$^{-1}$.  \\

Several simulations with different values for the scale length
$r_{\rm pl}^{\rm sc}$ of the super-cluster, different initial
masses $M_{\rm sc}$  of the super-cluster and different orbits
around the host galaxy are performed to study the evolution of
the resulting merger object.  The computational setup closely
follows that used in previous studies (Fellhauer et al.\ 2002;
Fellhauer \& Kroupa 2002). \\

It is possible to follow the evolution with a particle-mesh
code that neglects dynamical effects of two-body relaxation,
because the half-mass (bulk) two-body relaxation times of the
single star clusters, which can be estimated by (Binney \&
Tremaine 1987) 
\begin{eqnarray}
  \label{eq:relax}
  t_{\rm relax} & = & \frac{664}{\ln(0.5N)} \left( \frac{M_{\rm
        cl}} {10^{5}{\rm M}_{\odot}} \right)^{1/2} \left(
    \frac{1{\rm M}_{\odot}} {\overline{m}} \right) \left(
    \frac{r_{0.5}} {1 {\rm pc}} \right)^{3/2} {\rm Myr},
\end{eqnarray}
is $\approx 800$~Myr for a $10^{4}$~M$_{\odot}$ star cluster
ranging up to $4.4$~Gyr for a $10^{6}$~M$_{\odot}$ star cluster,
while the merging timescale is much shorter (Fellhauer et al.\
2002).  Furthermore, as shown below, the resulting merger objects
have relaxation times of a Hubble-time or longer.

\section{Results}
\label{sec:results}

\begin{figure}[t!]
  \begin{center}
    \epsfxsize=8cm
    \epsfysize=8cm
    \epsffile{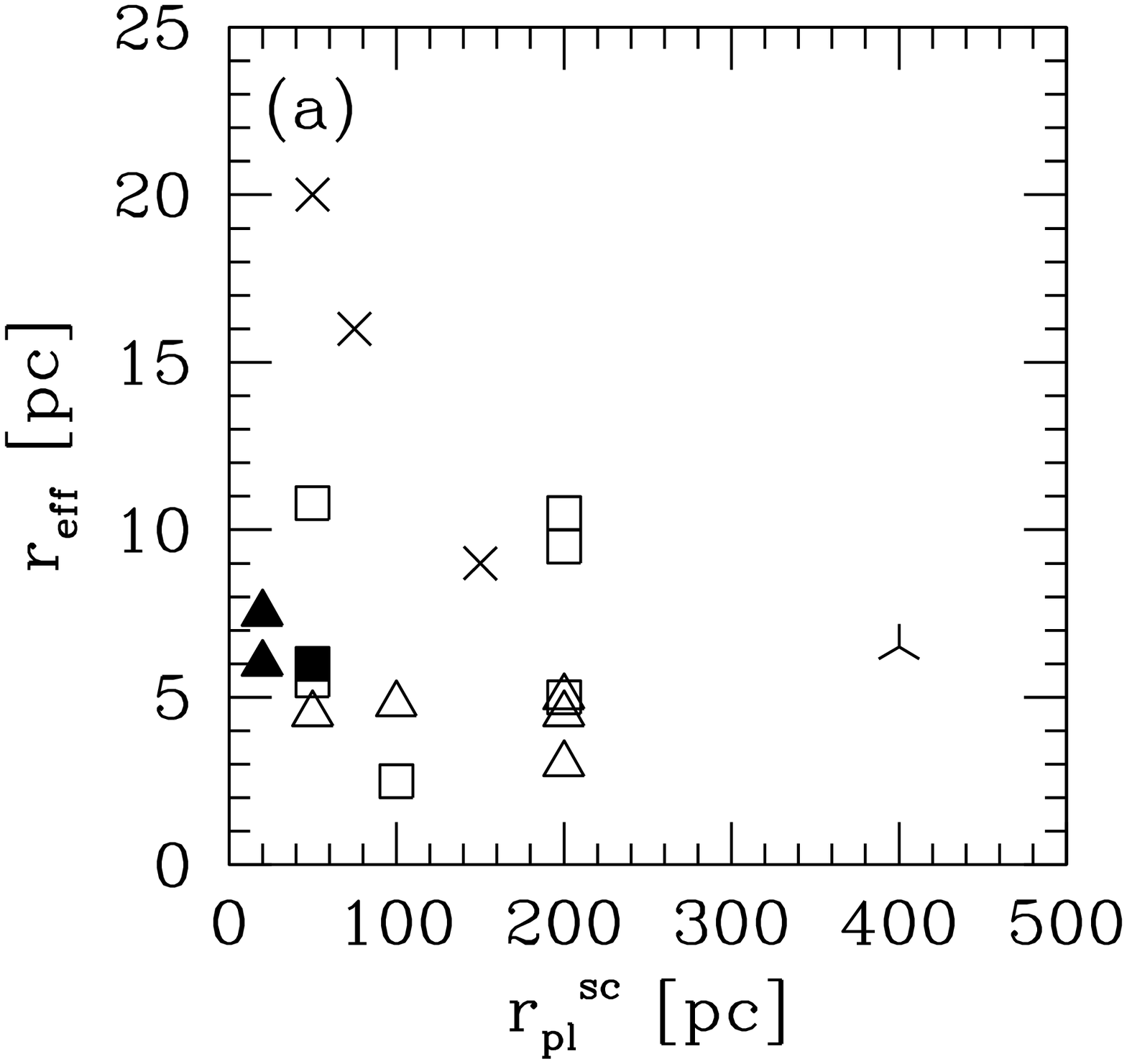}
    \epsfxsize=8cm
    \epsfysize=8cm
    \epsffile{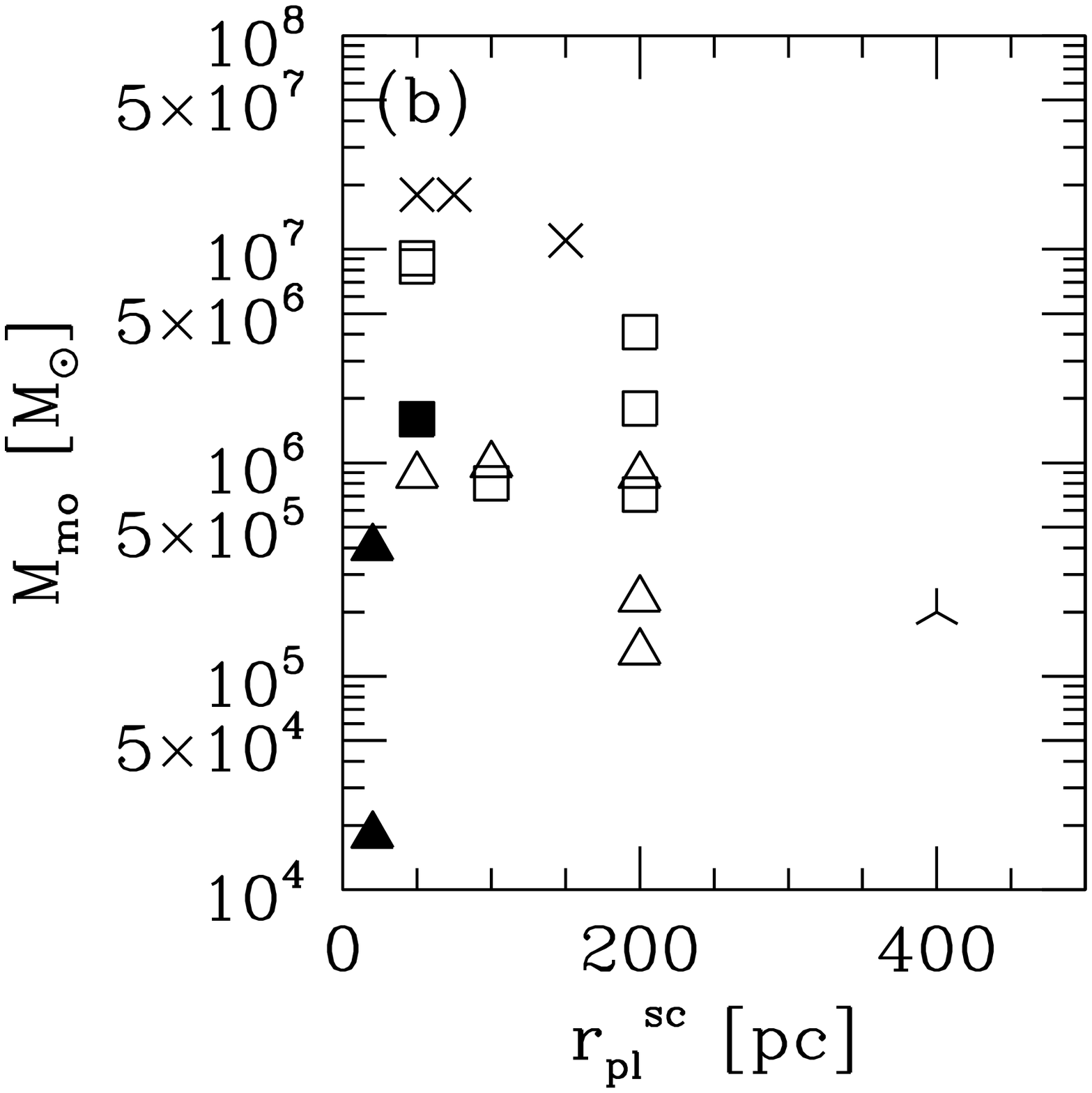}
    \caption{Effective radii, $r_{\rm eff}$, (a) and masses,
      $M_{\rm mo}$, (b) of our merger objects evaluated at
      $5$~Gyr vs.\ Plummer radius of the initial super-cluster
      $r_{\rm pl}^{\rm sc}$. Crosses: 
      large initial super-cluster mass ($M_{\rm sc} \geq
      10^{7}$~M$_{\odot}$) and weak tidal field (distance at
      perigalacticon $R_{\rm peri} \geq 10$~kpc); boxes: large
      mass and strong tidal field (distance at apogalacticon
      $R_{\rm apo} < 10$~kpc); triangles: low mass ($M_{\rm sc} <
      10^{7}$~M$_{\odot}$) and strong tidal field; tri-pointed
      stars: low mass and weak tidal field.  Filled symbols show
      the models discussed in the paper.} 
    \label{fig:reff}
  \end{center}
\end{figure}

In a previous paper (Fellhauer et al.\ 2002) it was shown that
the star clusters inside a super-cluster merge within a few
crossing times of the super-cluster (typically $50$ to
$500$~Myr).  The resulting merger objects are stable and compact.
Their masses range from being similar to GCs, in cases where the
tidal field removes a large fraction of the individual clusters
leaving merger objects consisting of only a few star clusters.
On the other hand, when the super-cluster is sufficiently
concentrated and the tidal field is correspondingly weak the
merger object evolves to become an ultra-compact dwarf galaxy
(UCD) (Fellhauer \& Kroupa 2002), a new class of objects
discovered recently in the Fornax galaxy cluster (Phillipps et
al.\ 2002; Hilker et al.\ 1999).  To show the different results
of a wide range of our simulations we plot the effective radii
and the masses of the merger objects as a function of the Plummer
radius of the super-cluster (Fig.~\ref{fig:reff}).  Even if tidal
forces are strong, compact and massive super-clusters evolve into
extended objects with lifetimes significantly longer than $5$~Gyr
although they orbit well inside the disc. \\  

Here we give the results of a simulation (OC03, filled box in
Fig.~\ref{fig:reff}) where the super-cluster was placed on an
orbit within the galactic disc which accounts for the scenario in
NGC~1023.  The resulting merger object is far too massive to be
one of the ``faint fuzzies'' found in NGC~1023, rather resembling
an object like $\omega$-Cen in our Milky Way but with two
additional simulations we show that these results still hold for
less massive objects like the faint extended star clusters in
NGC~1023. \\  

The super-cluster has initially a Plummer radius of $r_{\rm
  pl}^{\rm sc} = 50$~pc and a cut-off radius of $250$~pc.  It
contains 262 star clusters following the power law of
Eq.~\ref{eq:massspec} with a total mass of $10^{7}$~M$_{\odot}$.
The crossing time of the super-cluster is $t_{\rm cr}^{\rm sc} =
10.4$~Myr and the internal velocity dispersion of the clusters in
the super-cluster is $\sigma_{\rm sc} = 15.9$~kms$^{-1}$.
Additionally we choose the sense of rotation of all clusters in
the super-cluster to be the same to investigate the resulting
rotation-law of the merger object.  A super-cluster is expected
to rotate if it forms from a contracting and locally
differentially rotating inner tidal arm. The super-cluster is
placed on an eccentric orbit with perigalacticon at $2.1$~kpc and
apogalacticon at $7.5$~kpc.  The orbit is inclined such that the
maximum $Z$-distance from the disc plane is about $2$~kpc. The
parameters are chosen to be representative of the knots seen to
contain many star clusters in the Antennae galaxies (Fellhauer \&
Kroupa 2002), while the orbital inclination is motivated by the
orbit of $\omega$-Cen (Dinescu et al.\ 1999).  A possible link
between $\omega$-Cen and merged super-clusters has already been
pointed out by Fellhauer \& Kroupa (2002). \\ 

\begin{figure}[t!]
  \begin{center}
    \epsfxsize=8cm
    \epsfysize=8cm
    \epsffile{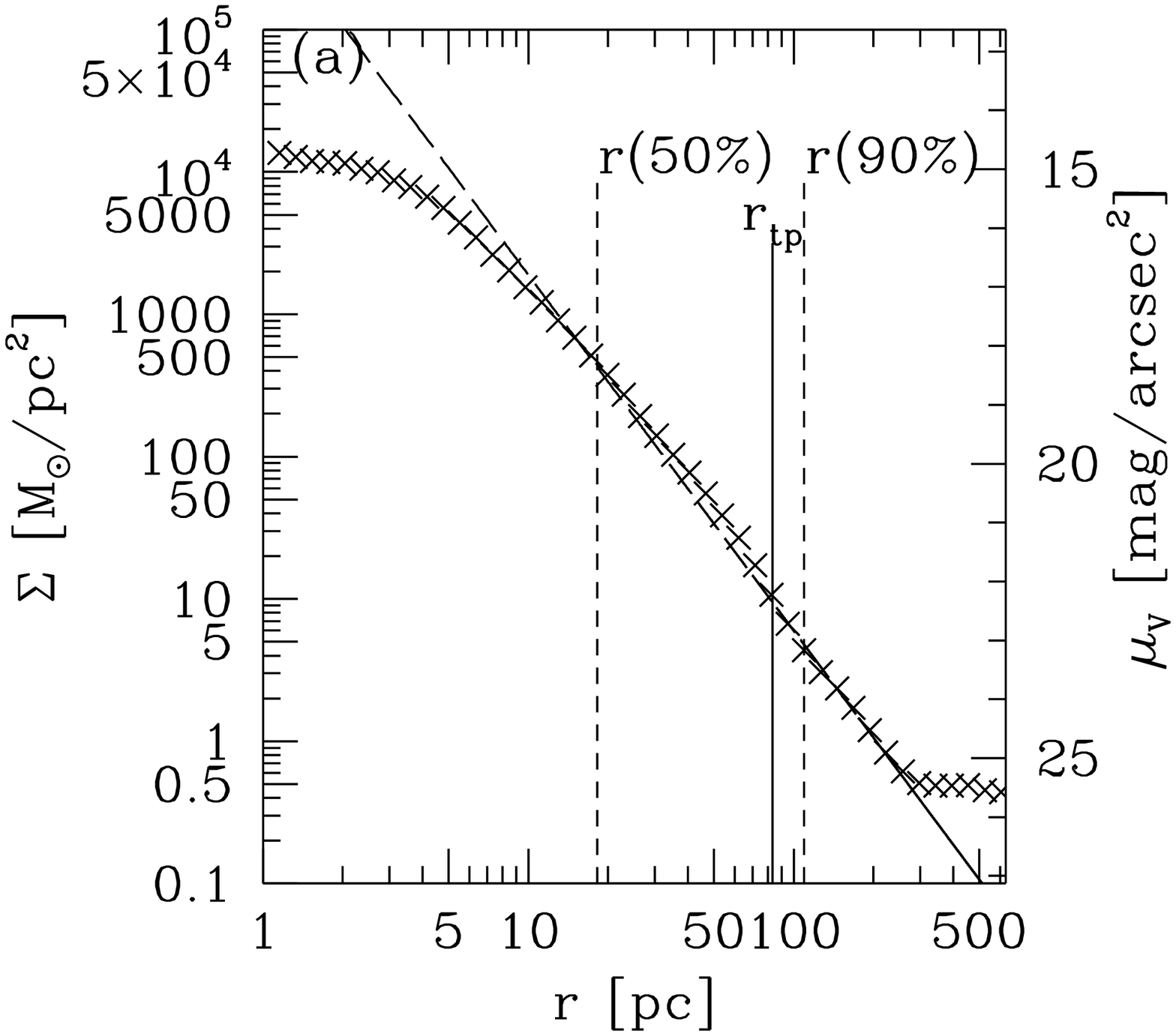}
    \epsfxsize=8cm
    \epsfysize=8cm
    \epsffile{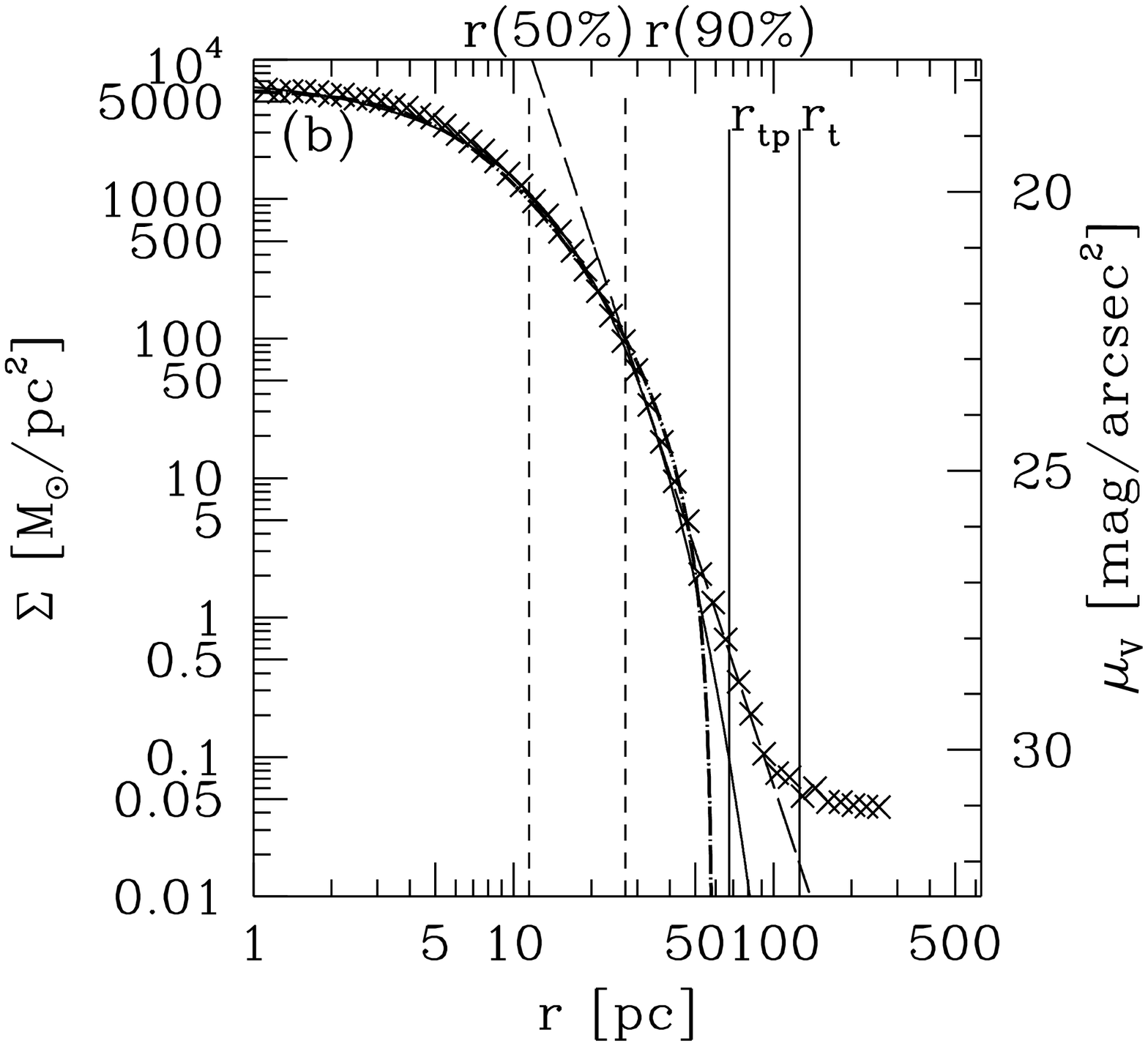}
    \caption{ Projected surface density profile of the merger
      object.  The vertical lines are as follows: dashed:
      half-mass and $90$\%-Lagrangian radius; solid: tidal radius
      at the position of the merger object ($r_{t}$) and at the
      last perigalacticon ($r_{t,p}$).  (a): Measured at $t =
      350$~Myr (shortly after formation of the merger object).
      The dashed line is a power law with $\Sigma \propto
      r^{-2.5}$.  The adopted mass-to-light ratio to calculate
      the surface-brightness is $0.3$.  (b): Profile at $t =
      5$~Gyr. Dot-dashed line shows the fitted King profile,
      solid line is the exponential fit for the inner part and
      dashed line is the fitted power law for the outer part.
      The adopted mass-to-light ratio is $3$.}  
    \label{fig:surf}
  \end{center}
\end{figure}

\begin{figure}[t!]
  \begin{center}
    \epsfxsize=8cm
    \epsfysize=8cm
    \epsffile{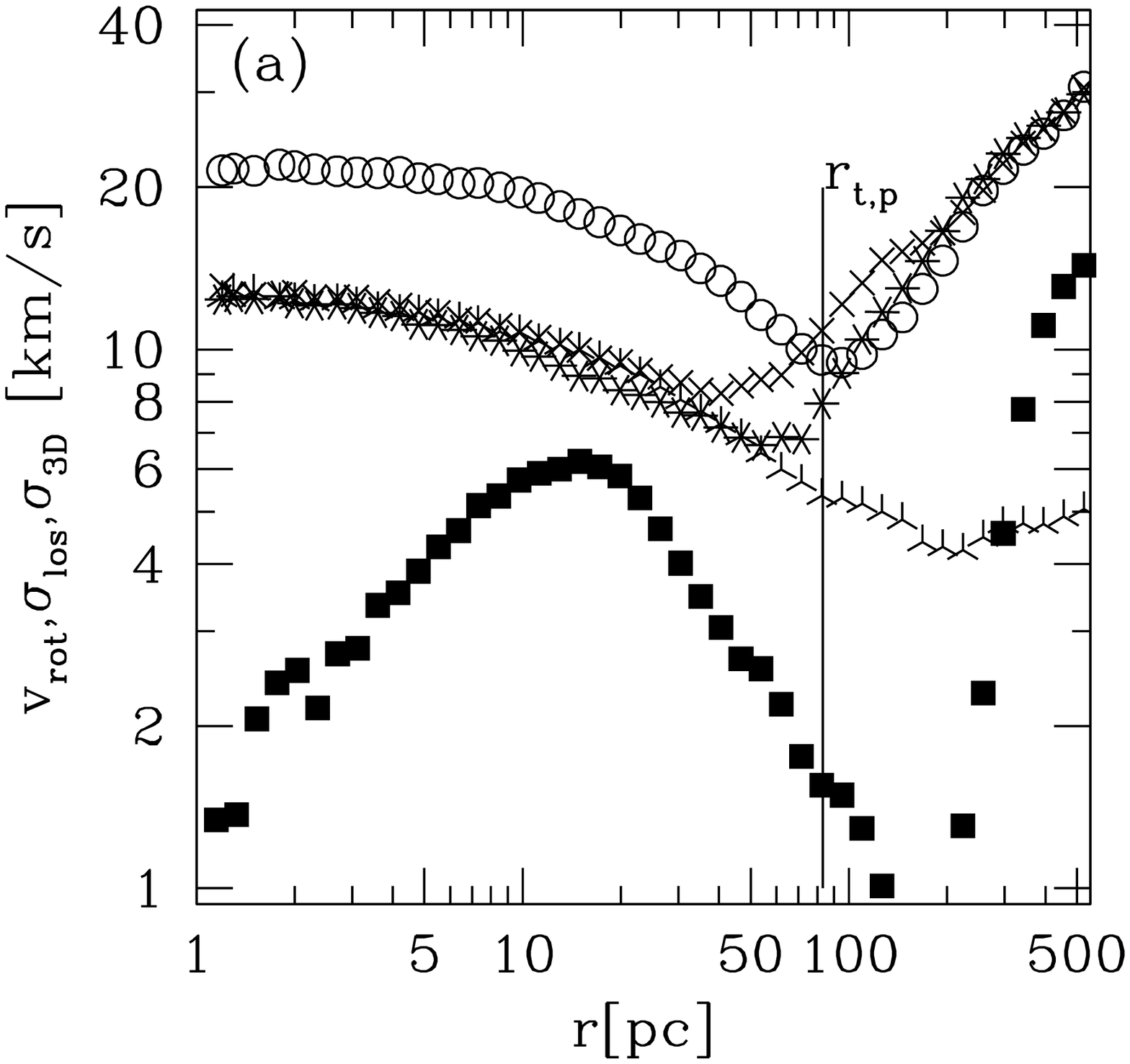}
    \epsfxsize=8cm
    \epsfysize=8cm
    \epsffile{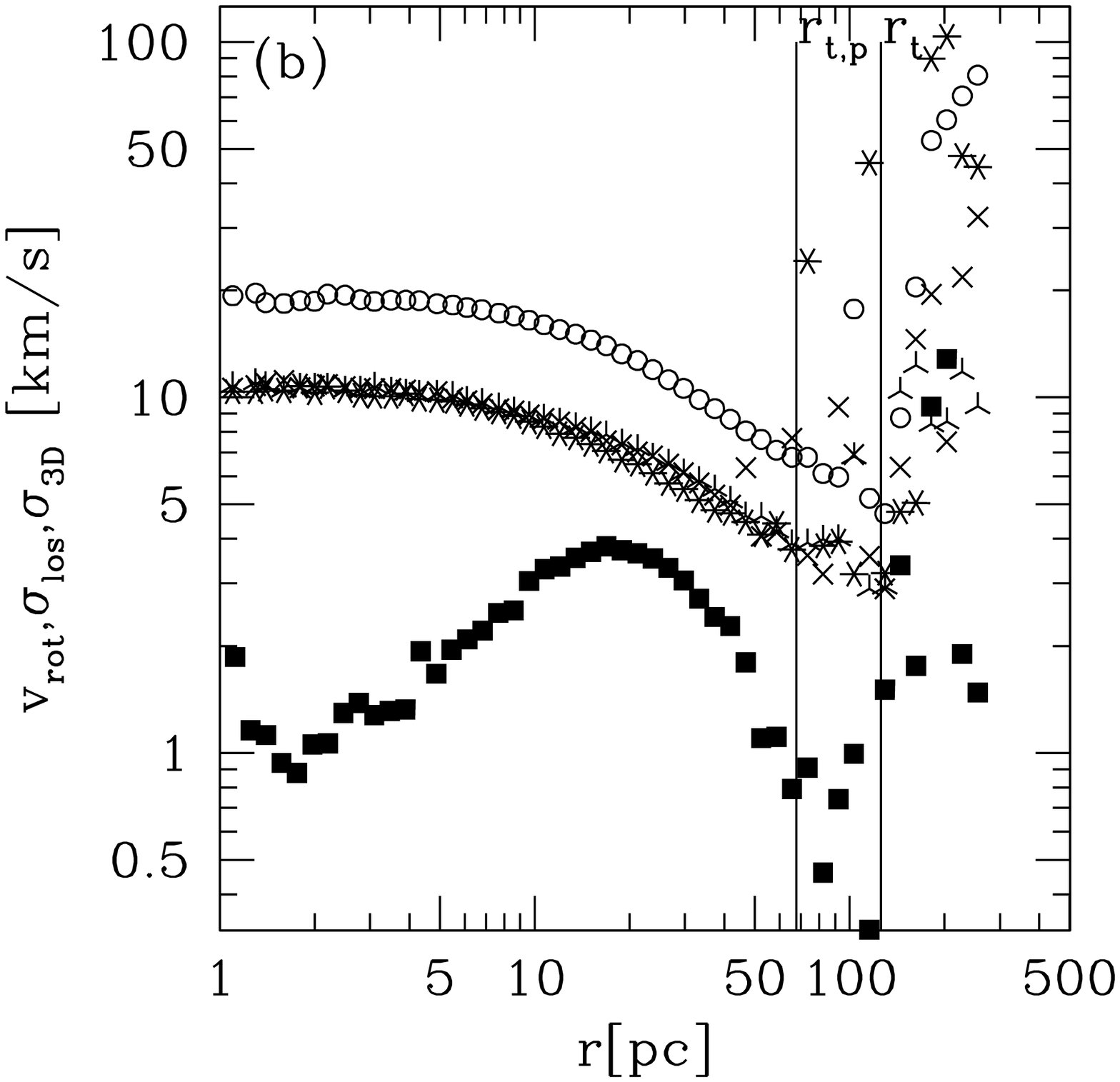}
    \caption{Velocity distribution in the merger object; circles
      show the 3-dimensional velocity dispersion computed in
      concentric shells around the centre; three-, four- and
      six-pointed stars are the projected line-of-sight velocity
      dispersions along the $x$-, $y$- and $z$-coordinate axis
      evaluated in concentric rings.  Squares show the mean
      rotational velocity again measured in concentric shells.
      Note the similarities to the rotation-law observed for
      $\omega$-Cen (Freeman 2001).  Vertical line shows the tidal
      radius at perigalacticon.  (a): Measured at $t = 350$~Myr
      (shortly after formation of the merger object).  (b):
      Distributions at $t = 5$~Gyr.}   
    \label{fig:vel}
  \end{center}
\end{figure}

Within the first $250$~Myr a merger object forms with a bound
mass of about $3 \cdot 10^{6}$~M$_{\odot}$.  The half-mass radius
of this object is $18$~pc and the $90$\% Lagrangian-radius is of
the order of $110$~pc at that time.  The surface density profile
of the merger object follows a shallow power law of $\Sigma
\propto r^{-2.5}$ as shown in Fig.~\ref{fig:surf}a. The
3-dimensional central velocity dispersion is $\sigma_{3D} =
21.8$~kms$^{-1}$ (Fig.~\ref{fig:vel}a). It also shows a maximum
rotation velocity of about $5$~kms$^{-1}$ at a distance of about
$12$~pc from the centre (Fig.~\ref{fig:vel}a).  The bulk
relaxation time (Eq.~\ref{eq:relax}) of the merger object is
$\approx 50$~Gyr which is longer than the age of the universe. \\

\begin{figure}[t!]
  \begin{center}
    \epsfxsize=8cm
    \epsfysize=8cm
    \epsffile{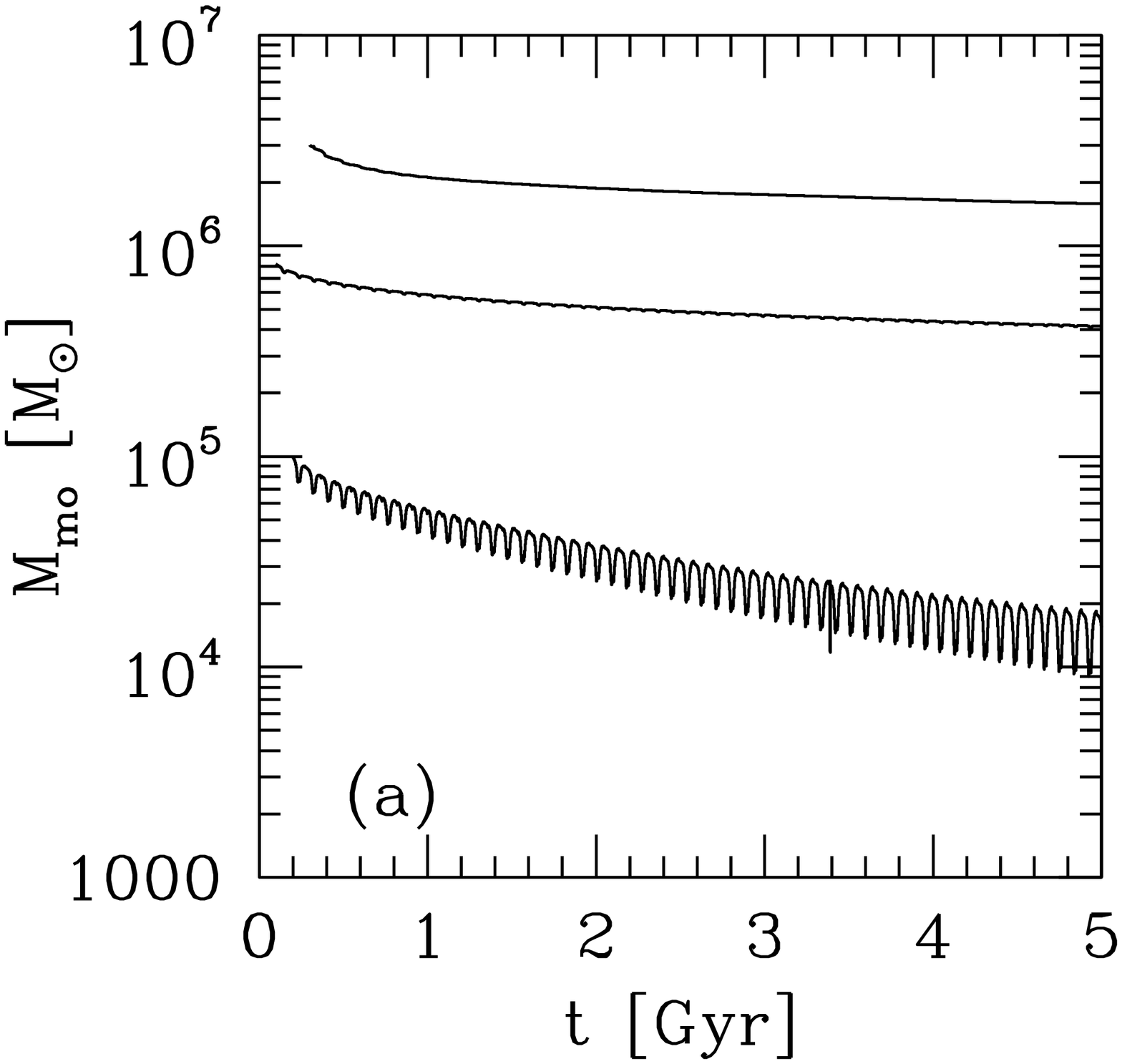}
    \epsfxsize=8cm
    \epsfysize=8cm
    \epsffile{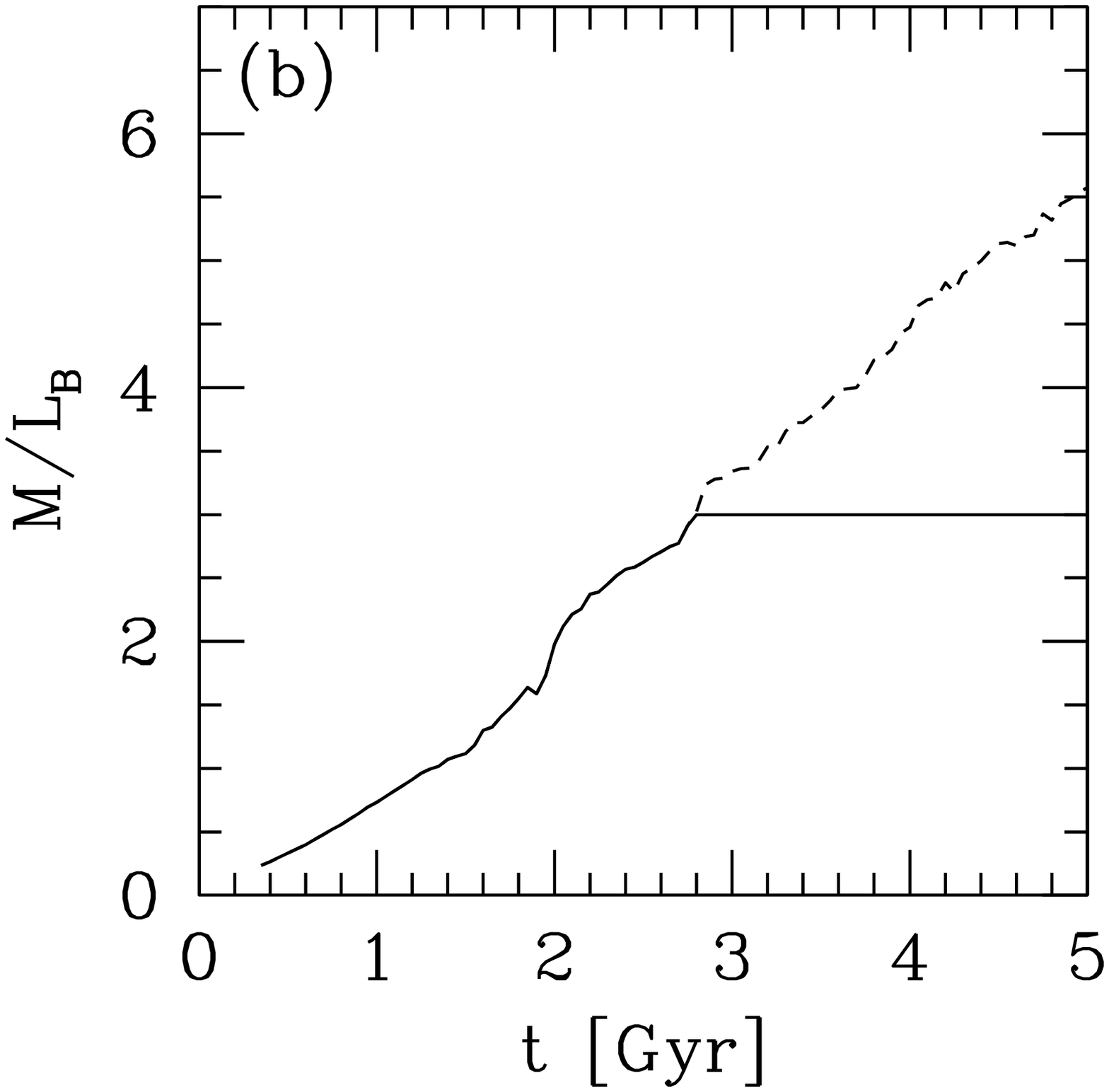}
    \caption{(a): Mass of the merger objects vs.\ time. From top
      to bottom: $M_{\rm sc} = 10^{7}$~M$_{\odot}$ (OC03),
      $M_{\rm sc} = 10^{6}$~M$_{\odot}$ (OC08) and $M_{\rm sc} =
      2 \cdot 10^{5}$~M$_{\odot}$ (OC07).  (b): Evolution of the
      mass-to-light ratio $M/L_{B}$ taken from {\sc Starburst99}
      (dashed line) and the ``frozen'' value ($M/L_{B} = 3.0$)
      used to calculate the luminosities (solid line).} 
    \label{fig:mass}
  \end{center}
\end{figure}

The tidal radius calculated using the equation 7-84 from Binney
\& Tremaine (1987)
\begin{eqnarray}
  \label{eq:rtidal}
  r_{\rm tidal}(D) & = & \left( \frac{M_{\rm mo}} {3M_{\rm
        gal}(D)} \right)^{1/3} D,
\end{eqnarray}
where $D$ denotes the distance to centre of the galaxy and
$M_{\rm gal}(D)$ is the enclosed mass of the galaxy at that
distance and $M_{\rm mo}$ is the bound mass of the merger object,
is about $84$~pc at perigalacticon.  Therefore 
the object becomes tidally shaped and looses mass with every
perigalacticon passage.  After $5$~Gyr of simulation time the
merger object has lost about half its mass
(Fig.~\ref{fig:mass}a).  The central surface density drops down
to $7500$~M$_{\odot}$pc$^{-2}$.  The profile can be fitted with a
King profile with King radius ($=$ effective radius) of $6$~pc.
An alternative fit is an exponential profile in the inner part
($\Sigma \propto e^{-r/6 {\rm pc}}$) and a steep power law with
$\Sigma \propto r^{-5.5}$ for the outer part
(Fig.~\ref{fig:surf}b).  The half-mass radius has shrunk down to
$11.6$~pc, the $90$\% Lagrangian-radius to $27$~pc.  The tidal
radius is determined to be $68$~pc at the last perigalacticon and
$127$~pc at the actual position.  Also the central velocity
dispersion and the maximum rotation speed has dropped as shown in 
Fig.~\ref{fig:vel}b.  The bulk relaxation time (following
Eq.~\ref{eq:relax}) of the merger object is now $t_{\rm relax}
\approx 20$~Gyr. \\ 

The two additional simulations have $N_{0}=20$ star clusters
initially with masses of $10^{4}$~M$_{\odot}$ ($M_{\rm sc} = 2
\times 10^{5}$~M$_{\odot}$) (OC07) and $5 \times
10^{4}$~M$_{\odot}$ ($M_{\rm sc} = 10^{6}$~M$_{\odot}$) (OC08)
respectively.  The Plummer-radii of the super-clusters are chosen
to be $20$~pc.  The orbits of the super-clusters have the same
eccentricity as in the first simulation but are placed in the
$X$-$Y$-plane of the galaxy instead of being inclined. \\

\begin{figure}[t!]
  \begin{center}
    \epsfxsize=8cm
    \epsfysize=8cm
    \epsffile{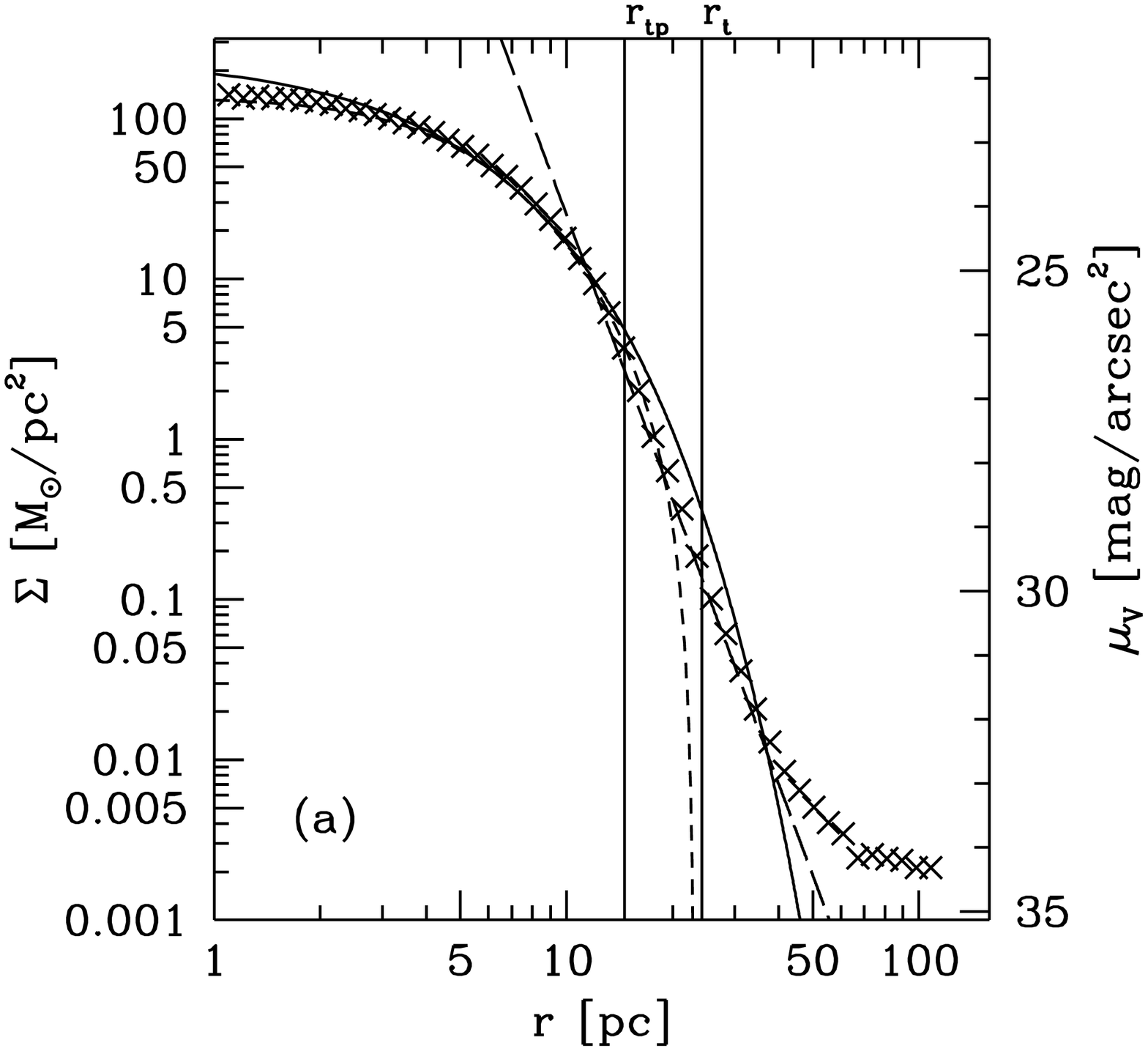}
    \epsfxsize=8cm
    \epsfysize=8cm
    \epsffile{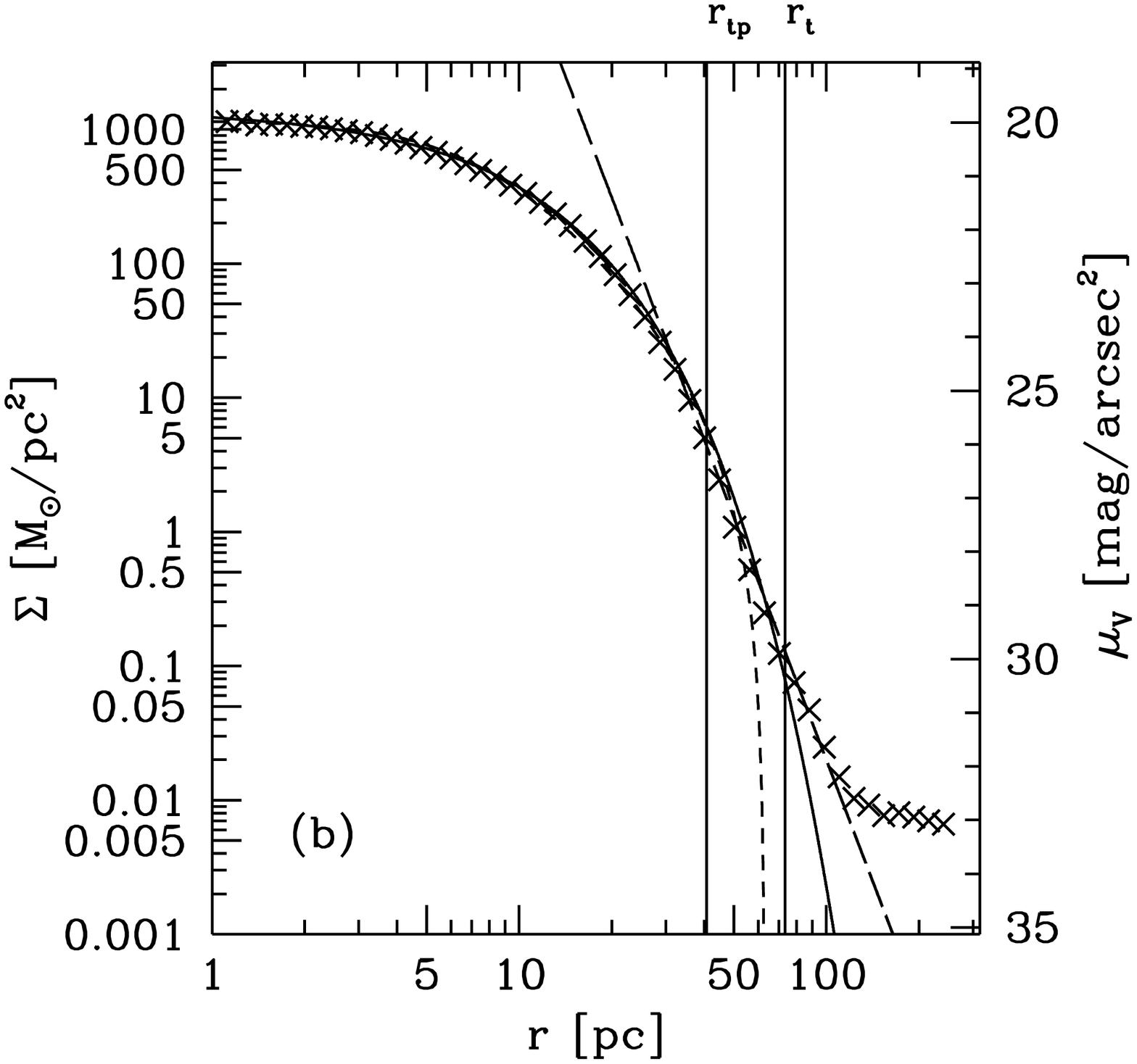}
    \caption{ Projected surface density profile of the merger
      objects.  The vertical lines are the calculated tidal radii
      at perigalacticon ($r_{t,p}$) and at the actual distance
      ($r_{t}$) (at $t=5$~Gyr) of the merger objects.  The solid
      curves show the fitted exponential profiles, the short
      dashed the fitted King profile and the long dashed the
      fitted power law with the values according to the main
      text.  (a) OC07: Simulation with $M_{\rm sc} = 2 \cdot
      10^{5}$~M$_{\odot}$; (b) OC08: Simulation with $M_{\rm sc}
      = 10^{6}$~M$_{\odot}$.}  
    \label{fig:surf2}
  \end{center}
\end{figure}

In the OC07 simulation a merger object containing $14$ out of the
initial $20$ star clusters develops after $100$~Myr.  After
$5$~Gyr this merger object has a bound mass of $1.8
\cdot 10^{4}$~M$_{\odot}$ (Fig.~\ref{fig:mass}a) and a tidal
radius of $24$~pc.  It's surface density profile can be fitted
with a King profile with a central surface density of
$240$~M$_{\odot}$pc$^{-2}$ and a King radius (which corresponds to
the effective radius) of $5.9$~pc.  Fitting an exponential
profile to the inner part gives $\Sigma_{\rm exp} =
250$~M$_{\odot}$pc$^{-2}$ and an exponential scale length of
about $4$~pc.  The outer part can again be fitted with a steep
power law $\Sigma \propto r^{-5.9}$ (Fig.~\ref{fig:surf2}a).
The central line-of-sight velocity dispersion is
$1.4$~kms$^{-1}$.  \\

The OC08 simulation shows a merger object which
developed out of $18$ star clusters.  Again the merging process
is finished after about $100$~Myr.  The mass of the merger object
after $5$~Gyr is $4 \cdot 10^{5}$~M$_{\odot}$
(Fig.~\ref{fig:mass}a) and the tidal radius is $70$~pc.  Fitting
a King profile to the surface density distribution
(Fig.~\ref{fig:surf2}b) gives the following values: central 
surface density of $1500$~M$_{\odot}$pc$^{-2}$ and an effective
radius of $7.5$~pc.  The exponential fit has the values
$1400$~M$_{\odot}$pc$^{-2}$ and $r_{\rm exp} = 7.5$~pc.  The
power law fit to the outer part has an power-law index of
$-6.0$. \\ 

\begin{figure}[t!]
  \begin{center}
    \epsfxsize=12cm
    \epsfysize=12cm
    \epsffile{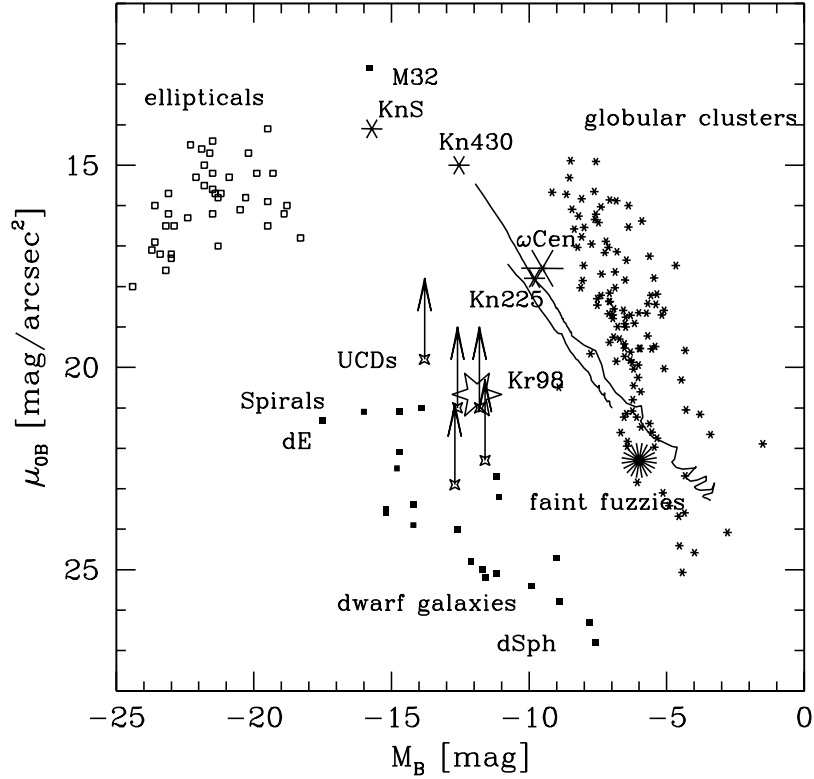}
    \caption{Central surface brightness, $\mu_{0,{\rm B}}$, vs
      absolute photometric B-band magnitude, $M_{\rm B}$ (the
      Kormendy diagram).  The stars are data for Milky-Way
      globular clusters (GC, Harris 1996), Local-Group dwarf
      galaxies (Mateo 1998) are displayed as filled boxes,
      elliptical galaxies (E, Peletier et al.\ 1990) are shown as
      open boxes.  The positions of disk galaxies are indicated
      by ``Spirals'' (cf. Ferguson \& Binggeli 1994).  The newly
      discovered UCDs are shown by arrows with lower-limits on
      $\mu_{\rm B}$ (Phillipps et al.\ 2001). Three ``knots'',
      (KnS, Kn430, Kn225, tables~1 and~2 in Whitmore et
      al. 1999), observed in the interacting Antennae galaxies
      are shown as stars.  Knots KnS and Kn430 are roughly 10~Myr
      old so that $M/L_{\rm B} \approx 0.035$, while Kn225 is
      about 500~Myr old ($M/L_{\rm B} \approx 0.476$).  The 
      open six-pointed star symbol shows the merged super-clusters
      predicted by Kroupa (1998) to result from the dynamical
      evolution of such knots, which consist of dozens to 
      hundreds of young star clusters.  The large fuzzy symbol
      shows the approximate location of the faint fuzzy star
      clusters.  The lines show the evolution of our merger
      objects from the time of formation until 5~Gyr using
      $M/L_{B}$ from Fig.~\ref{fig:mass}b and an additional
      linear fading for ages $> 3$~Gyr (from top-left to
      bottom-right: OC03, OC08 and OC07).}
    \label{fig:korm}
  \end{center}
\end{figure}

The above results (total mass and central surface density) can be
converted to total blue luminosity and central surface brightness
using mass-to-light ratios taken from a single stellar population
model of {\sc Starburst99} (Leitherer et al.\ 1999).  The
programme adopts a Salpeter IMF.  Therefore, we choose a minimum
stellar mass of $0.3$~M$_{\odot}$ and a maximum mass of
$100$~M$_{\odot}$.  Note that we truncate the Salpeter IMF at
$0.3$~M$_{\odot}$.  This roughly accounts for the flattening of
the IMF below $0.5$~M$_{\odot}$ (Kroupa 2002).  The evolution of
$M/L_{B}$ with time is shown in Fig.~\ref{fig:mass}b as obtained
as a result of stellar evolution from {\sc Starburst99}.  We stop
the photometric evolution from {\sc Starburst99} when $M/L_{B} =
3.0$ is reached because {\sc Starburst99} overestimates the $M/L$
at high ages quite substantially.  Instead we take a linear fading
with time as proposed by the models of Bruzual \& Charlot (1993)
of $2$ magnitudes from $3$ to $15$~Gyr.  This
enables us to place the evolutionary track of our merger objects
into a central surface brightness, $\mu_{0,{\rm B}}$, vs absolute
photometric B-band magnitude, $M_{\rm B}$, diagram (the Kormendy
diagram; Fig.~\ref{fig:korm}).  The endpoint of the evolution of
our merger objects is also the region where the ``faint fuzzies''
are located.  The faint fuzzies have total luminosities of about
$M_{V} = -7$ (Larsen \& Brodie 2002).  We estimate their
location in the Kormendy diagram in the following way.  First we
set $M_{V} + 1.0 = M_{B}$ (i.e.\ adopting $(B-V) = 1.0$ as an
admissible approximation for the diagram) then we convert the
total luminosity into a total mass using $M/L = 3.0$.  With this
mass we calculate the tidal radius in our model potential and
orbit ($r_{\rm tidal} \approx 50$~pc) and adopt an effective (or
King-) radius of $10$~pc.  Using a King profile we estimate the
central surface brightness to be $\mu_{0,B} \approx 22.3$~mag.  \\

\section{Conclusions}
\label{sec:conclus}

S0 (lenticular) galaxies are commonly believed to have suffered
from interactions with other galaxies which transformed their
morphology from a late-type spiral into their early-type form
(Kennicutt 1998; Abraham \& van den Bergh 2001).  Therefore the
type of NGC~1023 tells us that this galaxy suffered maybe several
but at least one strong interaction (merger) with another galaxy.
As is seen in interacting galaxies these events lead to an
enhanced star formation activity and to an enhanced rate of super
nova explosions of massive stars which accounts for the
super-solar enhancement of $\alpha$-elements.  Many of the new
stars mostly form in star clusters which are also grouped in
super-clusters, as is demonstrated by the Antennae and Stephan's
Quintet (Gallagher et al.\ 2001). These super-clusters are bound
entities and form, through successive mergers of their
constituent star clusters, new objects which are larger than
ordinary GCs.  Our simulations show that even strong tidal fields
are not able to destroy these objects and they survive for a
Hubble-time.  \\  

The ``faint fuzzies'' found in NGC~1023 may have therefore formed
as super-clusters during the interactions and the resulting star 
bursts that shaped NGC~1023.  Like in the Antennae the
super-clusters were born within the galaxy and not in the outer
tidal tails.  Therefore the ``faint fuzzies'', which one can call
extended globular clusters or small ultra-compact dwarf galaxies
by virtue of their long two-body relaxation time, stay within the
optically visible galaxy and show a similar rotation pattern like
the disc.  These objects can be old because they are able to
survive until the present, they may be rotating and have
different populations of stars if old field stars were captured
by the young super-cluster. \\ 

Given the ubiquity of young super-clusters in the Antennae and
Stephan's Quintet, it appears reasonable to expect that such
objects form whenever strong star bursts occur.  It is therefore
exciting, but retrospectively not very surprising, that ``faint
fuzzies'' were discovered in a lenticular galaxy. \\

\begin{figure}[t!]
  \begin{center}
    \epsfxsize=12cm
    \epsfysize=08cm
    \epsffile{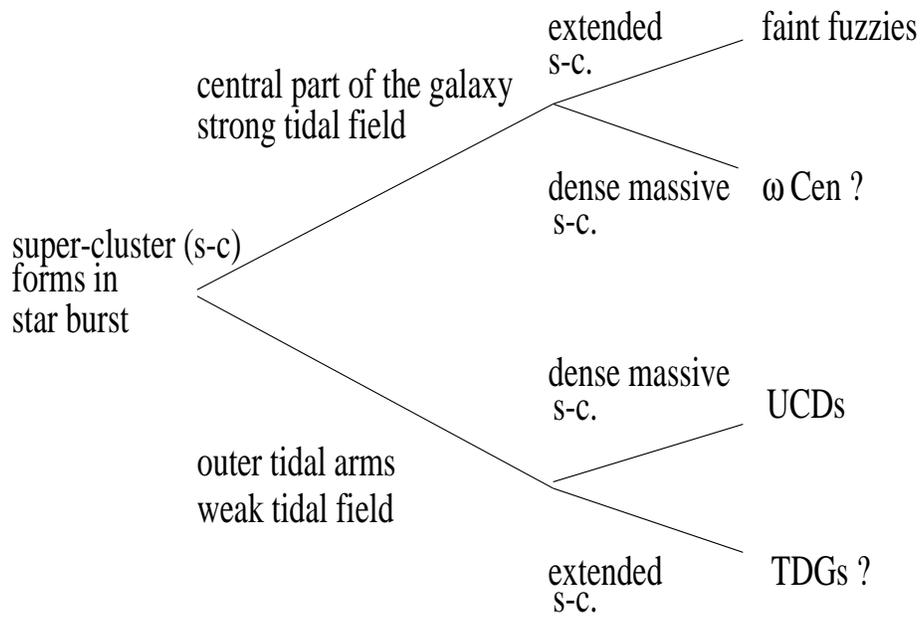}
    \caption{Schematic of possible evolutionary tracks of
      super-clusters.  Question marks signify on-going research.} 
    \label{fig:evol}
  \end{center}
\end{figure}

Super-clusters that form in outer tidal tails may leave the
mother galaxy either by being ejected by the merging
galaxies or become unbound due to a tidal field from the galaxy
cluster.  Such merger objects will appear similar to the UCDs in
Fornax (Fig.~\ref{fig:evol}).  A related object may be
$\omega$-Cen in our Milky Way, which is quite different and more
massive than the other GCs.  It incorporates at least two
different populations of stars and shows a rotation-law
resembling that of our merger object (Freeman 2001).
$\omega$-Cen is quite old and could therefore be a relic of a
star-burst during the formation of the Milky Way.  We are
investigating the hypothesis that $\omega$-Cen may also be
related to super-clusters.  Fig.~\ref{fig:evol} summarises in a
schematic way the possible evolutionary paths and resulting
objects that may form under this hypothesis.

\acknowledgements
MF acknowledges financial support through DFG-grant FE564/1-1.
MF also acknowledges the support of the European
Commission through grant number HPRI-1999-CT-00026 (the TRACS
programme at EPCC).

\clearpage

\end{document}